\documentclass[runningheads]{cl2emult}

\usepackage{makeidx}  
\usepackage{graphicx} 
\usepackage{subeqnar} 
\usepackage{multicol} 
\usepackage{cropmark} 
\usepackage{eso}      
\makeindex            

\def\ref{\par\noindent\hangindent=1truecm}

\catcode`\@=11
\def\gsim{\ifmmode{\mathrel{\mathpalette\@versim>}}
    \else{$\mathrel{\mathpalette\@versim>}$}\fi}
\def\lsim{\ifmmode{\mathrel{\mathpalette\@versim<}}
    \else{$\mathrel{\mathpalette\@versim<}$}\fi}
\def\@versim#1#2{\lower 2.9truept \vbox{\baselineskip 0pt \lineskip 
    0.5truept \ialign{$\m@th#1\hfil##\hfil$\crcr#2\crcr\sim\crcr}}}
\catcode`\@=12

\def\lsun{\hbox{$L_\odot$}}

\def\msun{\hbox{$M_\odot$}}

\def\ho{\hbox{$H_\circ$}}
\def\h50{\hbox{$\ho /50$}}
\def\yr-1{\hbox{${\rm yr}^{-1}$}}
\def\ha{\hbox{H$\alpha$}}
\def\hb{\hbox{H$\beta$}}

\def\cii{\hbox{C~{\scriptsize II}]}}
\def\ni{\hbox{[N~{\scriptsize I}]}}
\def\nii{\hbox{[N~{\scriptsize II}]}}
\def\sii{\hbox{[S~{\scriptsize II}]}}
\def\oi{\hbox{[O~{\scriptsize I}]}}
\def\oii{\hbox{[O~{\scriptsize II}]}}
\def\oiii{\hbox{[O~{\scriptsize III}]}}

\begin{document}
\title*{Supermassive Black Holes Can Hardly Be ``Silent''}
\toctitle{Supermassive Black Holes Can Hardly Be ``Silent''}
%
%
\titlerunning{No Silent Supermassive Black Holes}
%
\author{Alvio Renzini}
\authorrunning{A. Renzini}
%
%
\institute{European Southern Observatory, Garching b. M\"u nchen, Germany}

\maketitle              

\begin{abstract}
There is now ample evidence that most - perhaps all - galactic
spheroids host a supermassive BH at their center. This has been
assessed using a variety of observational techniques, from stellar
and/or gas dynamics to megamasers. Yet another promising technique is
offered by the case of the Virgo elliptical NGC 4552, in which early
HST/FOC observations revealed a central low-luminosity flare.
Subsequent HST/FOS observations with a 0.21 arcsec aperture have
revealed a rich emission-line spectrum with broad and narrow
components with FWHM of 3000 and 700 km/s, respectively. This
variable, mini-AGN at the center of NGC 4552 is most naturally the result of a sporadic accretion event on a central BH. It has a H$\alpha$ luminosity of only
$\sim 10,000\;\lsun$, making it the likely, intrinsically faintest AGN
known today. Only thanks to the superior resolution of HST such a
faint object has been discovered and studied in detail, but adaptive
optics systems on large ground-based telescopes may reveal in the
future that a low level of accretion onto central massive BHs is an
ubiquitous phenomenon among galactic spheroids.  FOC/FOS observations
of a central spike in NGC 2681 reveal several analogies  with the case 
of NGC 4552, while yet another example is offered by a recent
exciting finding with STIS by R.W. O'Connell in NGC 1399, 
the third galaxy in our original program. 
\index{abstract}


\end{abstract}

\section{Discovery of a Central Flare in NGC 4552}

Sophisticated techniques have been used to infer the  presence of
supermassive black holes (BH) at the center of galactic spheroids:
from
detailed stellar dynamical modeling fitting 2D spectroscopy data cubes
to megamaser observations, all aspects widely discussed at this meeting.
Here I will present observational evidence concerning three randomly-selected
 galactic
spheroids, showing that high spatial resolution imaging and/or spectroscopy
can reveal sporadic mini-AGN activity likely related to the presence
of a central  BH.

To investigate the UV upturn of ellipticals, in 1993  HST/FOC images in 
several UV bands were obtained for the
central regions of the elliptical galaxies NGC~1399 and NGC~4552 and of
the bulge of the S0/a galaxy NGC~2681. A point-like source -- or 
{\it spike} -- was evident at the center of both NGC~4552 and
NGC~2681, with their photometric profile being indistinguishable from
the PSF of the pre-COSTAR HST (Paper I [1]).

Comparison with another FOC image of NGC~4552 taken in 1991 [2]
showed that this spike had increased its
luminosity in the $U$ band (F342W) by a factor $\sim 7\pm1.5$ between
the two epochs, reaching $\sim 10^6\lsun$ (Paper I). A second point-like 
source is also present in the
1991 image, $\sim 0''.14$ from the central spike and with nearly the
same luminosity. While both sources were detected at the $\sim 4\sigma$
level in the 1991 image, the offcenter source was not detectable in the
1993 image at the $\sim 2.5\sigma$ level.

In Paper I  several possible interpretations were discussed for the
origin of this  {\it flare} at the center of NGC~4552,
favoring an accretion event onto a central massive black hole (BH). The
accreted material could have been tidally stripped from a star in a
close fly-by with the BH, though alternative ways of feeding the BH were
also considered. The low luminosity of the spike was nevertheless at variance
with the predicted luminosity in the case of a total stellar
disruption [3,4], hence the case of partial 
stripping was favored,  being also such
an event much more frequent than total disruptions.
 
\section{More HST Follow Up Reveals a Mini AGN in \\ NGC 4552}
If the flaring spike was due to accretion onto a BH, its spectrum
should show prominent, rotationally broadened emission lines, typical
of an accretion disk.  To test this expectation further HST
observations were obtained in 1996, both in imaging with FOC and in
spectroscopy with FOS.  The results of
these later HST observations are extensively reported in Paper II [5].

\subsection{FOC Imaging}
The 1991, 1993, and 1996 observations were made with different
telescope- instrument configurations and therefore differences in
detector efficiency and non-linearity effects were carefully treated
along with an adequate modeling of the PSF.  The analysis revealed
that the center spike in NGC~4552 increased its $U$-band luminosity by
a factor of 4.5 from 1991 to 1993 and faded from 1993 to 1996 by  a factor 
$\sim 2$ at all observed wavelengths (1700--3500 \AA)  (Paper
II).  In 1996 the UV fluxes indicated a black body temperature of
$T\sim15,000$ K (assuming the emission to be thermal) implying  a spike
bolometric luminosity of $\sim3\times10^5$ \lsun\  (at a
distance of 15.3 Mpc).  The offcenter spike that was present in 1991
did not appear in any of the later images.

\begin{figure}
\centering
\includegraphics[width=0.7\textwidth]{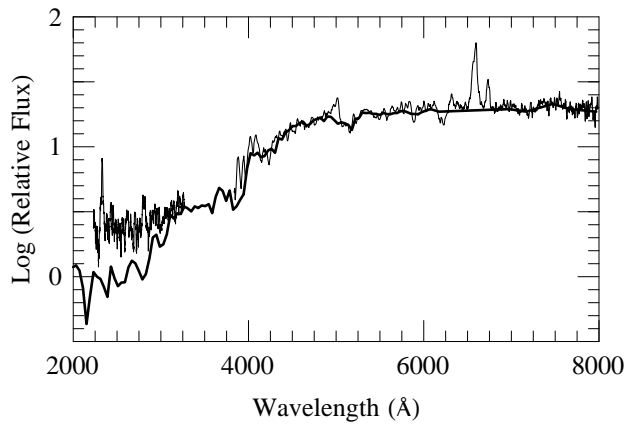}
\caption[]{The overall 1996 FOS spectrum of NGC~4552 within the
$0''.21\times0''.21$ aperture centered on the spike (thin  line), is
superimposed to a scaled combination of the IUE
$10''\times20''$\ aperture of NGC~4552 [6]
matched to ground-based optical spectrum of NGC~4649, a giant elliptical
whose SED is virtually the same as that of NGC~4552 [7]
(thick line).  The spectra have been normalized to the visual region. The
FOS spectrum show a strong UV excess and prominent emission lines. 
}
\label{eps1}
\end{figure}

\subsection{FOS Spectroscopy}
The 1996 FOS spectra were obtained through a $0''.21\times0''.21$
square aperture centered on the spike and covering the range from
$\sim 2200$ to $\sim 8500$ \AA.  Fig. 1 compares the 1996 FOS spectrum
of this central region with a composite spectrum meant to represent
the spectrum of the inner $r\le 7''$ of this galaxy.  This composite
spectrum is a good match to the SED of
the FOS spectrum, with two notable exceptions: 1. the FOS
spectrum shows strong emission lines that are absent in the composite
spectrum, and 2. shortward of 3200 \AA \  the FOS
spectrum is far stronger than the IUE SED.

The most prominent emission lines include \cii\ $\lambda$2326, MgII
$\lambda$2800, \oii\ $\lambda$3727, \sii\ $\lambda$4072, \hb, \oiii\
$\lambda\lambda$4959, 5007, \ni\ $\lambda$5700, \oi\ $\lambda$6300,
\nii\ $\lambda\lambda$6548, 6583, \ha\ and \sii\ $\lambda\lambda$6717,
6731.  The emission line  ratios of the narrow components
 place the spike among extreme AGNs, while the \oiii/\hb\
ratio falls just on the borderline between Seyferts and LINERs.

The emission line profiles were fitted with gaussian components,
reaching the following main  conclusions: 1. Good  fits
of the emission lines can be obtained only with a
combination of broad and narrow components for {\it both} the
permitted as well as the forbidden lines.  2. The emission lines are
very broad, with  very high velocity widths for both
the broad (FWHM $\simeq 3000$ km s$^{-1}$) and narrow components (FWHM
$\simeq 700$ km s$^{-1}$). 3. The shape of the \ha+\nii\ complex
has definitely changed from the 1996 spectrum to a FOS spectrum taken
8 months later, indicating a shift to the blue of $\sim 230$~km
s$^{-1}$ of the whole (narrow $+$ broad) H$\alpha$ line.

\begin{figure}
\centering
\includegraphics[width=0.7\textwidth]{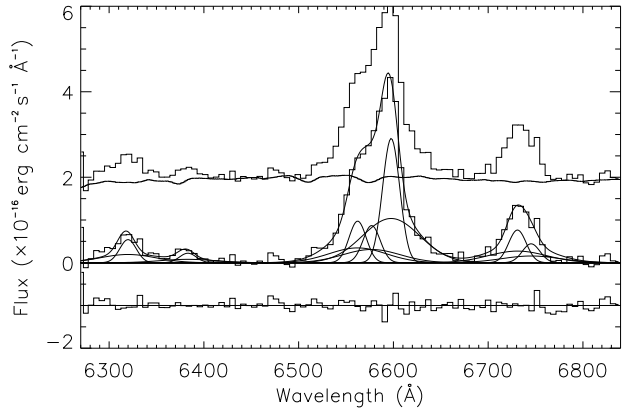}
\caption[]{The observed 
\oi, \nii, \ha, and \sii\ lines in the red region
of the FOS G780H spectrum (top) together with their gaussian decomposition 
into narrow and broad components (middle) and the corresponding residuals 
(bottom).
The narrow and broad components have FWHM$\simeq 700$ and $\simeq
 3000$ km s$^{-1}$, respectively.  }
\vskip -3 truemm
\label{eps2}
\end{figure}

\subsection{Interpretation}

This complex phenomenology clearly points towards the presence of a very
weak AGN at the center of this galaxy, most likely powered by a low
level of accretion onto a central BH. Other interpretations
were already considered
unlikely, given the early evidence (Paper I). The
additional evidence gathered in 1996 was conclusive in
this respect.
The mass of the supermassive BH
at the center of NGC~4552 was then estimated 
to be between $3\times10^8$ to $2
\times 10^9$ \msun, consistent with other ground-based estimates as well as
with the  BH mass-bulge mass (or -bulge luminosity) relation [8,9].

With the adopted distance  of 15.3 Mpc  the
broad \ha\
luminosity is $\sim 5.6\times 10^{37}$ erg s$^{-1}$, a factor of
two less than the broad \ha\ luminosity of the (previously) faintest
 known AGN, the
Seyfert 1 nucleus of NGC~4395, and a $\sim 20$ times fainter 
than M81, the next faintest Seyfert 1  nucleus [10].

As suggested in Paper I, this phenomenology 
is generically consistent with a scenario in which the
flare is caused by the tidal stripping of a star in a close
flyby with a central supermassive BH. From the theoretical side,
only the extreme case of the total disruption of a $\sim 1\,\msun$
main sequence star has been widely investigated so far (e.g., [3,4,11]).
The frequency of such events is estimated to be of one 
every $\sim 10^3-10^4$ yr in a giant elliptical galaxy [3].
The flare is predicted to be very bright for several years
($\sim 10^{10}\lsun$),  much brighter than the observed flare. This
indicates that if the flare in NGC~4552 was caused by a tidal stripping
in a BH-star flyby, then this flyby was rather wide and led to only
partial stripping.  One expects wider flybys to be vastly more
frequent than the hard ones causing total disruption.  To be consistent
with the observed luminosity, only $\sim 10^{-3}\msun$ should have been
stripped (Paper I), perhaps more if an ADAF is established (?).

Tidal stripping of a star is but one possible way to feed matter to a
massive central BH at a low rate.  Other mechanisms mentioned in Paper II
include: a) Roche
lobe overflow from a star in bound orbit around the BH; 
b) accretion from a clumpy interstellar medium which is actually seen
near the nucleus (see Fig. 12 in Paper II); or c) gas fed to the BH via a 
cooling flow within
the X-ray emitting hot interstellar medium.
Concerning alternative c), occasional cooling catastrophes 
lead to transient major cooling flow that can feed the central BH, while
mini-inflows may be active most of the time.
Such flows can lead to  sizable
excursions in the (mini-)AGN  luminosity (flickering), reminiscent of the
behavior of the spike in NGC 4552  [12,13].

It may well be that each  of these  mechanisms operates
from time to time in the central regions of elliptical galaxies like
NGC~4552, and in all these  options 
 an accretion disk is established, hence no 
clear cut discrimination can easily be achieved.
However, some evidence  favoring the tidal stripping option
comes from the noticed shift in the broad H$\alpha$ emission between
1996 and  1997. Tidal stripping/disruption is indeed predicted to
produce an {\it elliptical} accretion disk which precession
results in  H$\alpha$ line profile variations [14].
Finally, in Paper II it was
speculated that the offcenter spike seen only in the 1991 image could 
be due to a
relativistic jet emerging from the central mini-AGN having shocked the
dense dusty ring seen at nearly the same distance, with the jet
 having been possibly produced by a previous accretion event.

\section{Also NGC 2681 and NGC 1399 Display Central Activity}

In 1997 FOC and FOS observations were secured also for the central
spike in NGC 2681, with a virtually identical strategy to that used
for NGC 4552 (Paper III [15]). 
The main results can be summarized as follows.

 The photometric profile is well represented by a {\it Nuker-law} of
 the {\it power-law} type, from the innermost $0''.005$ up to
 $\sim100''$ from the center.  
Given the very high surface brightness of the central regions,
 a transient UV spike such as that in NGC 4552 would have not been
 revealed in either the FOC images (see Fig. 3) or in the FOS UV continuum.  
The UV continuum and the presence of
 Balmer absorptions indicate that the central region of NGC~2681 is
 dominated by the light of a relatively young stellar population (1--2
 Gyr). 

Contrary to the case of NGC 4552, the FOS spectrum 
of the innermost $0''.21\times0''.21$ region matches very well
the IUE UV spectrum, in spite of the much larger (a factor of
$\sim$4000) area sampled by the IUE aperture. Together with the
absence of gradients in the (Far-UV)$-$(Near-UV) and UV$-$IR colors,
this implies the homogeneity of the stellar population within $r\lsim
10''$.
The FOS shows 
emission lines whose ratios indicate that the nucleus is a
LINER. The emission lines are well-modeled by a single Gaussian with 
FWHM$\simeq480$ km s$^{-1}$, which is a
factor $\sim2$ higher than that measured from the ground, within a
$2''\times4''$ aperture, indicating the presence of a
central mass concentration.

This steepening of the (gas) velocity dispersion is not accounted for
by a spherical isotropic dynamical model with constant $M/L$, derived
by deprojecting the Nuker-law. The same kind of model gives a good fit
to the FOS data when assuming a central dark mass (BH) with
$M_\bullet\lsim6\times10^7$ \msun, consistent with the BH mass-bulge
luminosity relation [9].  This holds under the assumptions that the
emitting gas has an isotropic velocity-dispersion tensor and that its
density is proportional to the stellar density. Models without a BH
can also fit the data if these assumptions are relaxed, e.~g. either
the nuclear gas is in a disk, or gas clouds are on radially-anisotropic
orbits close to the nucleus. In summary, there are quite many
analogies between the central spikes in NGC 4552 and NGC 2681. The possibility
that the line emission in the NGC 2681 spike  originates in an
accretion disk seen more face on than in NGC 4552 cannot be ruled out.

\begin{figure}
\centering
\includegraphics[width=0.5\textwidth]{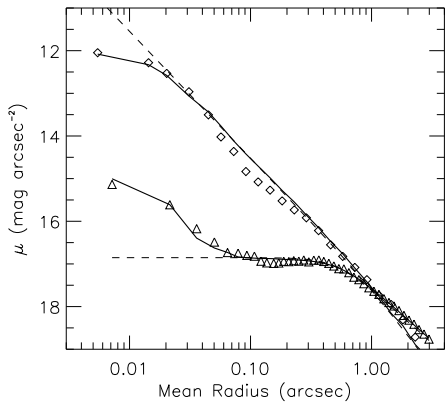}
\caption[]{The FOC $U$-band surface brightness profiles of the innermost
$3''$ of NGC 4552 (lower lines) and NGC 2681 (upper lines). The dashed
lines represent the best fit Nuker laws and the solid lines their
convolution with the PSF of the FOC, with a point-like source being
added in the case of NGC 4552.
}
\label{eps1}
\end{figure}

On 1993 FOC images the central $\sim 20''$ of NGC 1399 appear {\it as smooth
as silk}, i.e. no spike similar to either that of NGC 4552 or that of NGC 2681 is present. Indeed, NGC 1399 is well fit by a Nuker law of the core type, and
its central surface brightness is much lower than that of NGC 2681.
However, a STIS far UV spectrum of NGC 1399 recently obtained (R.W. O'Connell
private communication) appear to contain a distinct nuclear
point source --definitely above the background light of the galaxy's
core-- which may be even brighter than the spike in NGC 4552.
A very nice surprise indeed!

\section{No Supermassive Black Hole Can Be Totally Silent}

The case of the flare in NGC 4552 adds yet another option at our disposal
to gather evidence for supermassive BHs lurking at the center of apparently
{\it in-active} galactic nuclei.
Supermassive BHs in galactic spheroids  sit
at the bottom of their gravitational potential well, where
stellar and ISM densities reach their maximum, and where any
cannibalized material tends to converge.  As such, one is tempted to say
that the real problem is how to {\it avoid} a low, fluctuating level of
accretion onto a massive BH --- hence of low level AGN activity ---
rather than how to produce it. After all, it must be hard for guests as bulky
as $10^8$ or $10^9\;\msun$ to really hide at the center of galaxies:
wherever a massive
BH exists, its presence is likely to be betrayed by at least a low level of
AGN-like {\it activity}.

The case of NGC~4552 offers a lesson in this respect. Thanks to its
angular resolution, HST observations in either UV or optical imaging or
narrow aperture spectroscopy allow to reveal a mini-AGN activity which
would be essentially invisible  to similar ground based observations.
However, a  high resolution comparable  to that of HST
is now possible  also from the ground at near-IR
wavelength thanks to  adaptive optics (AO). AO-fed,
near-IR, 1D and 2D spectroscopy of the center of galactic
spheroids (e.g., with SINFONI at the VLT) 
might reveal broad emission lines of the Paschen and
Brackett series, hence potentially offering additional opportunities
for the demography of central BHs in galactic spheroids. The generic
prediction is that virtually all galactic spheroids (and especially
those for which dynamical evidence exists) should show a low-level AGN
activity similar to that so clearly detected by HST in NGC 4552.

\smallskip
\noindent
I would like to thank Bob O'Connell for his permission to mention his
discovery of a newly appeared point-like source in NGC 1399. It is also
a pleasure to thank the Paper I-III team, and especially Michele
Cappellari who made an effort to conclude the analysis of NGC 2681 in
time for this meeting.

\end{document}